\documentclass[pra,twocolumn,showpacs,superscriptaddress,floatfix,amsmath,nofootinbib,amssymb]{revtex4} 

\usepackage[english]{babel}
\usepackage{graphicx}
\usepackage{dcolumn}
\usepackage{bm}
\usepackage{verbatim}
\usepackage{color}

\newcommand{\ket}[1]{\left| {#1} \right\rangle}
\newcommand{\bra}[1]{\left\langle {#1} \right|}

\newcommand{\proj}[2]{\left| {#1} \right\rangle\!\left\langle {#2} \right|}

\newcommand{\dimctwo}{l}
\newcommand{\BbbR}{\mathbb{R}}

\usepackage{color,xspace}

\def\slashchar#1{\setbox0=\hbox{$#1$} 
\dimen0=\wd0 
\setbox1=\hbox{/} \dimen1=\wd1 
\ifdim\dimen0>\dimen1 
\rlap{\hbox to \dimen0{\hfil/\hfil}} 
#1 
\else 
\rlap{\hbox to \dimen1{\hfil$#1$\hfil}} 
/ 
\fi}

\begin{document}
\title{The Unruh effect  in quantum information beyond the single-mode approximation}

\author{David E. Bruschi}
\address{School of Mathematical Sciences, University of Nottingham, Nottingham NG7 2RD, United Kingdom}

\author{Jorma Louko}
\address{School of Mathematical Sciences, University of Nottingham, Nottingham NG7 2RD, United Kingdom}

\author{Eduardo Mart\'{i}n-Mart\'{i}nez}
\address{Instituto de F\'{i}sica Fundamental, CSIC, Serrano 113-B, 28006 Madrid, Spain}

\author{Andrzej Dragan}
\address{Institute of Theoretical Physics, University of Warsaw, Ho\.{z}a 69, 00-049 Warsaw, Poland}

\author{Ivette Fuentes\footnote{Previously known as Fuentes-Guridi and Fuentes-Schuller.}}
\address{School of Mathematical Sciences, University of Nottingham, Nottingham NG7 2RD, United Kingdom}

\begin{abstract}
We address the validity of the single-mode approximation that is commonly invoked in the analysis of entanglement in non-inertial frames and in other relativistic quantum information scenarios. We show that the single-mode approximation is not valid for arbitrary states, finding corrections to previous studies beyond such approximation in the bosonic and fermionic cases. We also exhibit a class of wave packets for which the single-mode approximation is justified subject to the peaking constraints set by an appropriate Fourier transform.
\end{abstract}

\maketitle

\section{Introduction}
The question of understanding entanglement in non-inertial frames has been central to the development of the emerging field of relativistic quantum information \cite{Alsingtelep,TeraUeda2,ShiYu,Alicefalls,AlsingSchul,SchExpandingspace,Adeschul,KBr,LingHeZ,ManSchullBlack,PanBlackHoles,AlsingMcmhMil,DH,Steeg,Edu2,schacross,Ditta,Hu,DiracDiscord,Edu6}. The main aim of this field is to incorporate relativistic effects to improve quantum information tasks (such as quantum teleportation) and to understand how such protocols would take place in curved space-times. In most quantum information protocols entanglement plays a prominent role. Therefore, it is of a great interest to understand how it can be degraded \cite{Alicefalls,AlsingSchul,Edu3,Edu4,Edu6} or created \cite{Ball,Expanfer,Edu7} by the presence of horizons or spacetime dynamics. 

Previous analyses show that entanglement between modes of bosonic or fermionic fields are degraded from the perspective of observers moving  in uniform acceleration. In this paper,  we analyze the validity of the single-mode approximation commonly used in such analyses and show that the approximation is justified only for a special family of states.  The single-mode approximation, which was introduced in \cite{Alsingtelep,AlsingMcmhMil} has been extensively used in the literature not only in discussions concerning entanglement but also in other relativistic quantum information scenarios \cite{Alicefalls,AlsingSchul,Bradler,highdim,chapucilla,chapucilla2,Edu2,Shapoor,matsako,Edu3,Ditta,Edu4,Geneferm,Edu5,DiracDiscord}. A deeper understanding of how the Unruh effect degrades entanglement is of crucial importance not only for fundamental questions but also to engineer a practical method to experimentally detect such effect. So far, not only has the effect not been measured but its very existence has been subject to some controversy (see, for instance, \cite{co1,co2,co3}).

In the canonical scenario considered in the study of entanglement in non-inertial frames the field, from the inertial perspective, is considered to be in a state where all modes are in the vacuum state except for two of them which are in a two-mode entangled state. For example, the Bell state
\begin{equation}\label{maxent1}
\ket{\Psi}_\text{M}=\frac{1}{\sqrt2}\left(\ket{0_{\omega}}_\text{M}\ket{0_{\omega^{\prime}}}_\text{M}+\ket{1_{\omega}}_\text{M}\ket{1_{\omega^{\prime}}}_\text{M}\right),
\end{equation}
where M labels Minkowski states and $\omega$, $\omega^{\prime}$ are two Minkowski frequencies.  Two inertial observers, Alice and Bob, each carrying a monocromatic detector sensitive to frequencies $\omega$ and $\omega^{\prime}$ respectively,  would find maximal correlations in their measurements since the Bell state is maximally entangled. It is then interesting to investigate to what degree the state is entangled when described by observers in uniform acceleration. In the simplest scenario,  Alice is again considered to be inertial and an uniformly accelerated observer Rob is introduced, who carries a monocromatic detector sensitive to mode $\omega^{\prime}$.  To study this situation, the states corresponding to Rob must be transformed into the appropriate basis, in this case, the Rindler basis.  It is then when the  single-mode approximation is invoked to relate Minkowski single particle states $\ket{1_{\omega^{\prime}}}_\text{M}$  to states in Rindler space. 

We argue that the single-mode approximation is not valid for general states. However, the approximation holds for a family of peaked Minkowski wave packets provided constraints imposed by an appropriate Fourier transform are satisfied.  We show that the state analyzed canonically in the literature corresponds to an entangled state between a Minkowski and a special type of Unruh mode. We therefore revise previous results for both bosonic and fermionic field entanglement.  The results are qualitatively similar to those obtained under the single-mode approximation.  We confirm that entanglement  is degraded with acceleration, vanishing in the infinite acceleration limit in the bosonic case and reaching a non-vanishing minimum for fermionic fields.  However, we find that in the fermionic case, the degree to which entanglement is degraded depends on the election of Unruh modes.

The paper is organized as follows:  we introduce in section \ref{sec3}  the transformations between Minkowski, Unruh and Rindler modes. In section \ref{sec:entganglementrev}, we analyze the entanglement degradation due to the Unruh effect for scalar fields including corrections to the single-mode approximation.  We exhibit in section \ref{sec:peaking} states for which the single-mode approximation is justified in the massless and massive bosonic case. In section \ref{sec5}  the degradation of entanglement between fermionic modes is addressed.  Finally, conclusions and discussions are presented in section \ref{conclusions}.

\section{Minkowski, Unruh and Rindler modes}\label{sec3}

We consider a 
real massless scalar field $\phi$ in a 
two-dimensional Minkowski spacetime. 
The field equation is the massless 
Klein-Gordon equation, $\Box \phi=0$. 
The (indefinite) 
Klein-Gordon inner product reads 
\begin{align}
(\phi_1,\phi_2) = i \int_{\Sigma} \phi_1^* \overleftrightarrow{\partial_a}\phi_2 \, n^a  \, \text{d}\Sigma,
\end{align}
where $n^a$ is a future-pointing normal vector to the spacelike hypersurface $\Sigma$ and $\text{d}\Sigma$ 
is the volume element on~$\Sigma$. 

The Klein-Gordon equation can be solved in Minkowski coordinates
$(t,x)$ which are an appropriate choice for inertial observers. The
positive energy mode solutions with respect to the timelike Killing
vector field $\partial_t$ are given by
\begin{align}
\label{eq:masslessMmodes}
u_{\omega,\text{M}} (t,x) = \frac{1}{\sqrt{4\pi \omega}}
\exp[-i\omega(t-\epsilon x)],
\end{align}
where $\omega>0$ is the Minkowski frequency and the discrete index $\epsilon$ 
takes the value $1$ for modes with positive momentum (the right-movers) 
and the value $-1$ for modes with negative momentum (the left-movers). 
As the right-movers and the left-movers decouple, we have suppressed the index $\epsilon$ on the left-hand side of \eqref{eq:masslessMmodes} and we continue to do so in all the formulas. The mode solutions and their complex conjugates are normalised in the usual sense of Dirac delta-functions in $\omega$ as 
\begin{eqnarray}
\left(u_{\omega,\text{M}},  u_{\omega',\text{M}}\right)& =& 
\delta_{\omega\omega'},\nonumber\\
\left(u^{\ast}_{\omega,\text{M}},  u^{\ast}_{\omega',\text{M}}\right)& =& 
- \delta_{\omega\omega'} ,\nonumber\\ 
\left(u^{\ast}_{\omega,\text{M}},  u_{\omega',\text{M}}\right)& =& 0.
\end{eqnarray}

The Klein-Gordon equation can also be separated in coordinates
that are adapted to the Rindler family of 
uniformly accelerated observers. 
Let region I (respectively region II) denote the wedge $|t|<x$ 
($x<-|t|$). In each of the wedges, we introduce the Rindler
coordinates $(\eta,\chi)$ by \cite{Takagi} 
\begin{equation}
\label{Rindlertransformation}
\eta = \text{atanh} \! \left(\frac{t}{x}\right), 
~~~\chi = \sqrt{x^2-t^2},
\end{equation}
\begin{figure}[h]
\begin{center}
\includegraphics[width=.50\textwidth]{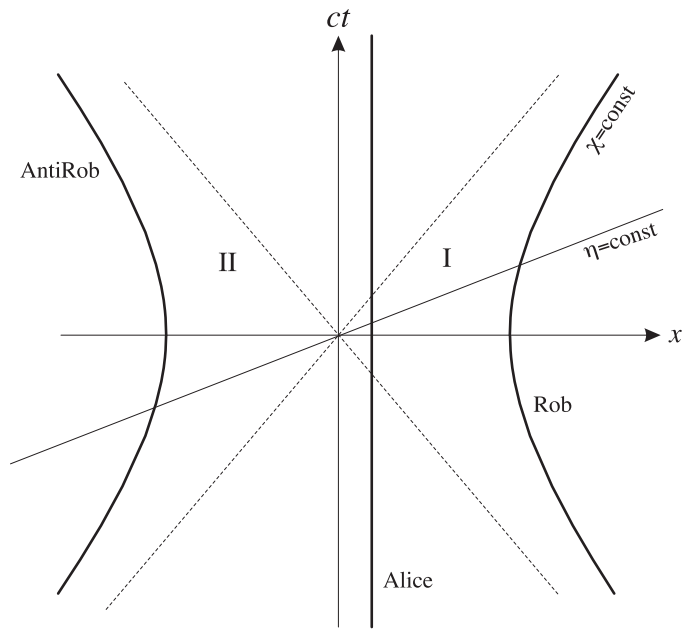}
\end{center}
\caption{ Rindler space-time diagram: lines of constant position $\chi=\text{const.}$ are hyperbolae and all curves of constant  $\eta$ are straight lines that come from the origin. An uniformly accelerated observer Rob travels along a hyperbola constrained to either region I or region II.}
\label{figbosons}
\end{figure}
where $0<\chi<\infty$ and $-\infty < \eta < \infty$ individually in each
wedge. The curve $\chi = 1/a$, where $a$ is a positive constant
of dimension inverse length, is then the world line of a
uniformly-accelerated observer whose proper acceleration equals~$a$,
and the proper time of this observer is given by $\eta/a$ in I and by
$-\eta/a$ in~II\null.  Note that $\partial_\eta$ is a timelike Killing
vector in both I and~II, and it is future-pointing in I but
past-pointing in~II\null.

Separating the Klein-Gordon equation in regions I and II in the
Rindler coordinates yields the solutions
\begin{eqnarray}
u_{\Omega,\text{I}} (t,x) &=& \frac{1}{\sqrt{4\pi \Omega}}
{\left(\frac{x-\epsilon t}{l_{\Omega}}\right)}^{i\epsilon\Omega},\nonumber\\
u_{\Omega,\text{II}} (t,x) &=& \frac{1}{\sqrt{4\pi \Omega}}
{\left(\frac{\epsilon t - x}{l_{\Omega}}\right)}^{-i\epsilon\Omega}, 
\end{eqnarray}
where $\epsilon = 1$ again corresponds to right-movers and $\epsilon =
-1$ to left-movers, 
$\Omega$ is a positive dimensionless constant 
and $l_{\Omega}$ is a positive constant of dimension length. 
As $\partial_\eta
u_{\Omega,\text{I}} = -i\Omega u_{\Omega,\text{I}}$ 
and 
$\partial_\eta 
u_{\Omega,\text{II}} = i\Omega u_{\Omega,\text{II}}$, 
$u_{\Omega,\text{I}}$ and $u_{\Omega,\text{II}}$ are the
positive frequency mode functions 
with respect to the future-pointing Rindler Killing
vectors $\pm \partial_\eta$ in their respective wedges, 
and $\Omega$ is the dimensionless Rindler frequency. 
The dimensional frequency with respect to the
proper time of a Rindler observer located at $\chi = 1/a$ is given
in terms of the dimensionless $\Omega$ by $\Omega_a = a\Omega$. The modes are 
delta-normalised in $\Omega$ in their respective
wedges as usual.

Note that the choice of the constant $l_{\Omega}$ is equivalent to
specifying the phase of the Rindler modes. This choice is hence purely
a matter of convention, and it can be made independently for each
$\Omega$ and~$\epsilon$. We shall shortly specify the choice so
that the transformation between the Minkowski and Rindler modes
becomes simple.


A third basis of interesting solutions to the field equation 
is provided by the Unruh modes, 
defined by 
\begin{eqnarray}
\label{eq:unruhmodes}
u_{\Omega,\text{\text{R}}} &=& 
\cosh(r_\Omega)  u_{\Omega,\text{I}} + \sinh(r_\Omega)  u^{\ast}_{\Omega,\text{II}}, \nonumber\\
u_{\Omega,\text{\text{L}}} &=& 
\cosh(r_\Omega)  u_{\Omega,\text{II}} + \sinh(r_\Omega) u^{\ast}_{\Omega,\text{I}},
\end{eqnarray} 
where $\tanh r_\Omega= e^{-\pi\Omega}$. 
While the Unruh modes have a sharp Rindler frequency, an analytic continuation argument shows that they are
purely positive frequency linear combinations 
of the Minkowski modes~\cite{Unruh,Birrell}. 
It is hence convenient to examine the transformation 
between the Minkowski and Rindler modes in two stages:  
\begin{itemize}
\item The well-known transformation 
\eqref{eq:unruhmodes} between the Unruh and Rindler modes isolates the consequences of the differing Minkowski and Rindler definitions of positive frequency.
\item The less well-known transformation between the Minkowski and Unruh modes \cite{Takagi} shows that a monochromatic wave in the Rindler basis  corresponds to a non-monochromatic superposition in the Minkowski basis.
\end{itemize}
 It is these latter effects from which the new observations in this paper will stem. 


To find the Bogoliubov transformations that relate the bases, 
we expand the field in each of the bases as 
\begin{eqnarray}
\!\!\phi &=&\!\! \int_0^\infty\left( a_{\omega,\text{M}} u_{\omega,\text{M}} + a_{\omega,\text{M}}^\dag u^{\ast}_{\omega,\text{M}} 
\right) \text{d}\omega\nonumber\\*
=& &\!\!\!\!\!\! \!\!\!\!\!\! \int_0^\infty \!\!\!\left( A_{\Omega,\text{\text{R}}} u_{\Omega,\text{\text{R}}}\!+\!A^{\dagger}_{\Omega,\text{\text{R}}} u^{\ast}_{\Omega,\text{\text{R}}}\!+\! A_{\Omega,\text{\text{L}}} u_{\Omega,\text{\text{L}}} \!+\! A^{\dagger}_{\Omega,\text{\text{L}}} u^{\ast}_{\Omega,\text{\text{L}}} \right)\! \text{d}\Omega\nonumber\\ \nonumber \\
=& &\!\!\!\!\!\! \!\!\!\!\!\!\int_0^\infty\!\!\!\left( a_{\Omega,\text{I}} u_{\Omega,\text{I}}\!+\!a^{\dagger}_{\Omega,\text{I}} u^{\ast}_{\Omega,\text{I}}\! +\! a_{\Omega,\text{II}} u_{\Omega,\text{II}}\!+\!a^{\dagger}_{\Omega,\text{II}} u^{\ast}_{\Omega,\text{II}}\right)\! \text{d}\Omega,
\label{eq:fieldexpansions}
\end{eqnarray}
where $a_{\omega,\text{M}}$, $A_{\Omega,\text{\text{R}}}, A_{\Omega,\text{\text{L}}}$, and $a_{\Omega,\text{I}}, a_{\Omega,\text{II}}$ are the Minkowski, Unruh and Rindler annihilation operators, respectively. The usual bosonic commutation relations $[a_{\omega,\text{M}},a^{\dagger}_{\omega^{\prime},\text{M}}] =\delta_{\omega\omega^{\prime}}$,  $[A_{\Omega,\text{\text{R}}},A^{\dagger}_{\Omega^{\prime},\text{\text{R}}}]=[A_{\Omega,\text{\text{L}}},A^{\dagger}_{\Omega^{\prime},\text{\text{L}}}]=\delta_{\Omega\Omega^{\prime}}$ and $[a_{\Omega,\text{I}},a^{\dagger}_{\Omega^{\prime},\text{I}}] = [a_{\Omega,\text{II}},a^{\dagger}_{\Omega^{\prime},\text{II}}]=\delta_{\Omega\Omega^{\prime}}$ hold, and commutators for mixed $\text{\text{R}}$, $\text{\text{L}}$ and mixed $\text{I}$, $\text{II}$ vanish. The transformation between the 
Unruh and Rindler bases is given by~\eqref{eq:unruhmodes}. The transformation between the Minkowski and Unruh bases can be evaluated by taking appropriate inner products of formula \eqref{eq:fieldexpansions} with the mode functions~\cite{Takagi}, 
with the result 
\begin{eqnarray}
u_{\omega,\text{M}} &=& \int_{0}^{\infty} \left(\alpha^\text{\text{R}}_{\omega\Omega} u_{\Omega,\text{\text{R}}}+  \alpha^\text{\text{L}}_{\omega\Omega} u_{\Omega,\text{\text{L}}}\right) \text{d}\Omega, \nonumber \\
u_{\Omega,\text{\text{R}}} &=& \int_{0}^{\infty} (\alpha^\text{\text{R}}_{\omega\Omega})^{\ast} u_{\omega,\text{M}} \, \text{d}\omega, \nonumber \\
u_{\Omega,\text{\text{L}}} &=& \int_{0}^{\infty}(\alpha^\text{\text{L}}_{\omega\Omega})^{\ast} u_{\omega,\text{M}} \, \text{d}\omega,
\label{eq:Mink-v-Unruh}
\end{eqnarray}
where 
\begin{eqnarray}
\label{eq:alphas-raw}
\alpha^\text{\text{R}}_{\omega\Omega} &=&\frac{1}{\sqrt{2\pi\omega}}\sqrt{\frac{\Omega\sinh\pi\Omega}{\pi}}\Gamma(-i\epsilon\Omega){(\omega l_{\Omega})}^{i\epsilon\Omega} , \nonumber \\
\alpha^\text{\text{L}}_{\omega\Omega} &=& \frac{1}{\sqrt{2\pi\omega}}\sqrt{\frac{\Omega\sinh\pi\Omega}{\pi}}\Gamma(i\epsilon\Omega){(\omega l_{\Omega})}^{-i\epsilon\Omega}.
\end{eqnarray}
By the properties of the Gamma-function 
(\cite{NISTlibrary}, formula 5.4.3), we can take 
advantage of the arbitrariness of the constants 
$l_{\Omega}$ and choose them so that 
\eqref{eq:alphas-raw} simplifies to 
\begin{eqnarray}
\alpha^\text{\text{R}}_{\omega\Omega} &=&\frac{1}{\sqrt{2\pi\omega}}{(\omega l)}^{i\epsilon\Omega} , \nonumber \\
\alpha^\text{\text{L}}_{\omega\Omega} &=& \frac{1}{\sqrt{2\pi\omega}}{(\omega l)}^{-i\epsilon\Omega} , 
\label{eq:alphas}
\end{eqnarray}
where $l$ is an overall constant of dimension length, 
independent of $\epsilon$ and~$\Omega$. 

The transformations between the modes give rise to transformations between the corresponding field operators. From~\eqref{eq:Mink-v-Unruh}, 
the Minkowski and Unruh operators are related by 
\begin{eqnarray} 
\label{eq:Aaboth-transform}
a_{\omega,\text{M}} &=& \int_0^{\infty}\left[(\alpha^\text{\text{R}}_{\omega\Omega})^{\ast} A_{\Omega,\text{\text{R}}}+(\alpha^\text{\text{L}}_{\omega\Omega})^{\ast} A_{\Omega,\text{\text{L}}}\right] \text{d}\Omega , \nonumber \\
A_{\Omega,\text{\text{R}}} &=& \int_0^{\infty} \alpha^\text{\text{R}}_{\omega\Omega} a_{\omega,\text{M}} \, \text{d}\omega , \nonumber \\
A_{\Omega,\text{\text{L}}} &=& \int_0^{\infty} \alpha^\text{\text{L}}_{\omega\Omega} \, a_{\omega,\text{M}} \, \text{d}\omega,
\end{eqnarray}
and from~\eqref{eq:unruhmodes}, the Unruh and Rindler operators 
are related by 
\begin{eqnarray}
a_{\Omega,\text{I}} &=&  \cosh(r_\Omega)\,A_{\Omega, \text{R}} + \sinh(r_\Omega)\, A^\dag_{\Omega,\text{\text{L}}} , \notag \\
a_{\Omega,\text{II}} &=& \cosh(r_\Omega)\, A_{\Omega,\text{\text{L}}} + \sinh(r_\Omega)\, A^\dag_{\Omega,\text{\text{R}}}.
\label{eq:Mink-v-Unruh-operators}
\end{eqnarray}

We can now investigate how the vacua and excited states 
defined with respect to the different bases are related. Since the transformation between the Minkowski and Unruh bases does not mix the creation and annihilation operators, these two bases share the common Minkowski vacuum state $|0\rangle_{\text{M}}=|0\rangle_\text{U}=\prod_{\Omega}|0_{\Omega}\rangle_\text{U}$, where $A_{\Omega,\text{\text{R}}}|0_{\Omega}\rangle_\text{U}=A_{\Omega,\text{\text{L}}}|0_{\Omega}\rangle_\text{U}=0$. However, $|0\rangle_\text{U}$ does not 
coincide with the Rindler vacuum: 
if one makes the ansatz 
\begin{align}
\label{eq:UOmegavac-expansion}
|0_{\Omega}\rangle_\text{U}= \sum_n f_{\Omega}(n) \, |n_{\Omega}\rangle_\text{I} |n_{\Omega}\rangle_\text{II}, 
\end{align}
where $|n_{\Omega}\rangle_\text{I}$
is the state with $n$ Rindler $\text{I}$-excitations 
over the Rindler $\text{I}$-vacuum $|0_{\Omega}\rangle_\text{I}$, and 
similarly 
$|n_{\Omega}\rangle_\text{II}$
is the state with $n$ Rindler $\text{II}$-excitations 
over the Rindler $\text{II}$-vacuum $|0_{\Omega}\rangle_\text{II}$,
use of \eqref{eq:Mink-v-Unruh-operators} shows that the 
coefficient functions are given by 
$f_{\Omega}(n) =\tanh^n \! r_{\Omega}/\cosh r_\Omega$. 
$|0\rangle_\text{U}$ is thus a two-mode squeezed state of Rindler excitations over the Rindler vacuum for each~$\Omega$.

Although states with a completely sharp value of 
$\Omega$ are not normalisable, 
we may approximate 
normalisable wave packets that are sufficiently narrowly peaked in 
$\Omega$ by taking a fixed $\Omega$ 
and renormalising the Unruh and 
Rindler commutators to read 
$[A_{\Omega,\text{\text{R}}},A^{\dagger}_{\Omega,\text{\text{R}}}]=[A_{\Omega,\text{\text{L}}},A^{\dagger}_{\Omega,\text{\text{L}}}]=1$ and 
$[a_{\Omega,\text{I}},a^{\dagger}_{\Omega,\text{I}}] = [a_{\Omega,\text{II}},a^{\dagger}_{\Omega,\text{II}}]=1$, with the commutators for mixed $\text{\text{R}}$, $\text{\text{L}}$ 
and mixed $\text{I}$, $\text{II}$ vanishing. 
In this idealisation of sharp peaking in~$\Omega$, 
the most general creation operator
that is of purely positive Minkowski frequency can be written as 
a linear combination of the two Unruh creation operators, in the form 
\begin{align}
\label{eq:q-defs}
a_{\Omega,\text{U}}^\dag = q_{\text{L}} A^\dag_{\Omega,\text{\text{L}}} + q_{\text{R}}  A^\dag_{\Omega,\text{\text{R}}},
\end{align}
where $q_{\text{R}}$ and $q_{\text{L}}$ are complex numbers with 
${|q_{\text{R}}|}^2 + {|q_{\text{L}}|}^2 =1$. Note that 
$[a_{\Omega,\text{U}}, a_{\Omega,\text{U}}^\dag]=1$. 
From 
\eqref{eq:UOmegavac-expansion} and \eqref{eq:q-defs}
we then see that adding into Minkowski vacuum one 
idealised particle
of this kind, of purely positive Minkowski frequency, yields the state 
\begin{align} 
& a^\dag_{\Omega,\text{U}} |0_\Omega\rangle_\text{U} =  
\sum_{n=0}^\infty 
f_{\Omega}(n)\frac{\sqrt{n+1}}{\cosh r_{\Omega}}|\Phi^{n}_{\Omega}\rangle
,  \notag 
\\
& |\Phi^n _{\Omega}\rangle = 
q_{\text{L}} \, |n_{\Omega}\rangle_\text{I}|(n+1)_{\Omega}\rangle_\text{II} +
q_{\text{R}} |(n+1)_{\Omega}\rangle_\text{I} |n_{\Omega}\rangle_\text{II} . 
\label{eq:singleOmegaex}
\end{align}

In previous studies on relativistic quantum information, it has been common to consider a state of the form \eqref{eq:singleOmegaex} with $q_{\text{R}}=1$ and $q_{\text{L}}=0$. The above discussion shows that this choice for $q_{\text{R}}$ and $q_{\text{L}}$ is rather special; in particular, it breaks the symmetry 
between the right and left Rindler wedges. We shall address next how entanglement is modified for these sharp $\Omega$ states when both $q_{\text{R}}$ and $q_{\text{L}}$ are present, 
and we then turn to examine the assumption of sharp~$\Omega$.

\section{Entanglement revised beyond the single mode approximation\label{sec:entganglementrev}}

In the relativistic quantum information literature, the single mode approximation $a_{\omega,\text{M}}\approx a_{\omega,\text{U}}$ is considered to relate Minkowski and Unruh modes. The main argument for taking this approximation is that the distribution
\begin{equation}
\label{eq:Aa-transform}
a_{\omega,\text{M}} = \int_0^{\infty}\left[
(\alpha^\text{\text{R}}_{\omega\Omega})^{\ast} A_{\Omega,\text{\text{R}}}+(\alpha^\text{\text{L}}_{\omega\Omega})^{\ast} A_{\Omega,\text{\text{L}}}\right]\text{d}\Omega
\end{equation}
is peaked. However,  we can see from equations \eqref{eq:alphas}  that this distribution in fact oscillates and it is not peaked at all.  Entanglement in non-inertial frames can be studied provided we consider the state
\begin{equation}\label{maxent}
\ket{\Psi}=\frac{1}{\sqrt2}\left(\ket{0_\omega}_{\text M}\ket{0_\Omega}_\text{U}+\ket{1_\omega}_{\text{M}}\ket{1_\Omega}_\text{U}\right),
\end{equation}
where the states corresponding to $\Omega$ are Unruh states. In this case, a single Unruh frequency $\Omega$ corresponds to the same Rindler frequency. In the special case  $q_\text{\text{R}}=1$ and  $q_{\text{L}}=0$ we recover the results canonically presented in the literature \cite{Alicefalls,AlsingSchul,Edu4}. In this section, we will revise the analysis of entanglement in non-inertial frames for the general Unruh modes. However, since a Minkowski monochromatic basis seems to be a natural choice for inertial observers, we will show in the section \ref{sec:peaking} that the standard results also hold for Minkowski states as long as special Minkowski wavepackets are considered. 

Having the expressions for the vacuum and single particle states in the  Minkowski, Unruh and Rindler basis enables us to return to the standard scenario for analyzing the degradation of entanglement from the perspective of observers in uniform acceleration. Let us consider the maximally entangled state Eq.~\eqref{maxent} from the perspective of inertial observers. By choosing different  $q_\text{\text{R}}$ we can vary the states under consideration. An arbitrary Unruh single particle state has different right and left components where $q_\text{\text{R}}, q_\text{\text{L}}$ represent the respective weighs. When working with Unruh modes, there is no particular reason why to choose a specific $q_\text{\text{R}}$. In fact, and as as we will see later, feasible elections of Minkowski states are in general, linear superpositions of different Unruh modes  with different values of $q_\text{\text{R}}$.

The Minkowski-Unruh state under consideration can be viewed as an entangled state of a  tri-partite system.  The partitions correspond to the three sets of modes:  Minkowski modes with  frequency $\omega$ and  two sets of Unruh modes (left and right) with frequency $\Omega$.  Therefore, it is convenient to define the following bi-partitions: the Alice-Bob bi-partition corresponds to Minkowski and right Unruh modes while the Alice-AntiBob bipartition refers to Minkowski and left Unruh modes. We will see that the distribution of entanglement in these bi-partitions becomes relevant when analyzing the entanglement content is the state from the non-inertial perspective. 

 We now want to study the entanglement in the state considering that the $\Omega$ modes are  described by observers in uniform acceleration.  Therefore, Unruh states must be transformed into the Rindler basis. The state in the Minkowski-Rinder basis is also a state of a tri-partite system.  Therefore, we define the Alice-Rob bi-partition as the Minkowski and region I Rindler modes while the Alice-AntiRob bi-partitions corresponds to Minkowski and region II Rindler modes.  In the limit of very small accelerations Alice-Rob and Alice-AntiRob approximate to Alice-Bob and Alice-AntiBob bi-partitions respectively. This is because, as shown in \eqref{eq:unruhmodes} and \eqref{eq:Mink-v-Unruh-operators},  region I (II) Rindler modes tend to R (L) Unruh modes in such limit.

The entanglement can be quantified using the Peres partial-transpose criterion. Since the partial transpose of a separable state has always positive eigenvalues, the a state is non-separable (and therefore, entangled) if the partial transposed density matrix has, at least, one negative eigenvalue. However, this is a sufficient and necessary condition only for $2\times 2$ and $2\times 3$ dimensional systems. In higher dimensions, the criterion is only necessary. Based on the Peres criterion a number of entanglement measures have been introduced. In our analysis we will use the negativity $\mathcal{N}$ to account for the quantum correlations between the different bipartitions of the system. It is defined as the sum of the negative eigenvalues of the partial transpose density matrix i.e., if $\lambda_\text{I}$ are the eigenvalues of any partially-transposed bi-partite density matrix $\rho_{AB}$ then its negativity is
\begin{equation}
\label{negativitydef}
\mathcal{N}_{AB}=\frac12\sum_{ i}(|\lambda_\text{I}|-\lambda_\text{I})=-\sum_{\lambda_\text{I}<0}\lambda_\text{I}.
\end{equation}
The maximum value of the negativity (reached for maximally entangled states) depends on the dimension of the maximally entangled state. Specifically, for qubits $\mathcal{N}_{AB}^{\text{max}}=1/2$.

In what follows we study the entanglement between the Alice-Rob and Alice-AntiRob modes. After expressing Rob's modes in the Rindler basis, the Alice-Rob density matrix is obtained by tracing over the region II, with the result
\begin{equation}
\rho_{AR} = \frac{1}{2}\sum_{n=0}^\infty {\left[\frac{\tanh^n r_\Omega}{\cosh r_\Omega}\right]}^2 \rho_{AR}^n,
\end{equation}
where 
\begin{eqnarray}
\nonumber\rho_{AR}^n&\!\!=&\!\!\proj{0n}{0n}+\frac{n+1}{\cosh^2r_{\Omega}}\Big(|q_\text{\text{R}}|^2\proj{1\,n+1}{1\,n+1}\\*
\nonumber&&+|q_\text{\text{L}}|^2\proj{1n}{1n}\Big)+\frac{\sqrt{n+1}}{\cosh r_{\Omega}}\Big(q_\text{\text{R}}\proj{1\,n+1}{0n}\\*
\nonumber&&+ q_\text{\text{L}}\tanh r_{\Omega}\proj{1\,n}{0\, n+1}\Big)+\frac{\sqrt{(n+1)(n+2)}}{\cosh^2r_{\Omega}}\\*
&&\times q_\text{\text{R}}\,q_\text{\text{L}}^*\tanh r_{\Omega}\proj{1\, n+2}{1n}+(\text{H.c.})_{_{\substack{\text{non-}\\\text{diag.}}}}.
\end{eqnarray}
Here $(\text{H.c.})_{\text{non-diag.}}$ means Hermitian conjugate of only the non-diagonal terms. The Alice-AntiRob density matrix is obtained by tracing over region I. However, due to the symmetry in the Unruh modes between region I and II, we can obtain the Alice-AntiRob  matrix by exchanging $q_\text{\text{R}}$ and $q_\text{\text{L}}$.  The partial transpose $\sigma_\text{\text{R}}$ of $\rho_\text{\text{R}}$ with respect 
to Alice is given by
\begin{equation}
\sigma_{AR} = 
\frac{1}{2}
\sum_{n=0}^\infty {\big[f(n)\big]}^2 \sigma^n_{AR}
\label{eq:sigmaR}
\end{equation}
where 
\begin{eqnarray}
\nonumber\sigma_{AR}^n&\!\!=&\!\!\proj{0n}{0n}+\frac{n+1}{\cosh^2r_{\Omega}}\Big(|q_\text{\text{R}}|^2\proj{1\,n+1}{1\,n+1}\\*
\nonumber&&+|q_\text{\text{L}}|^2\proj{1n}{1n}\Big)+\frac{\sqrt{n+1}}{\cosh r_{\Omega}}\Big(q_\text{\text{R}}\proj{0\,n+1}{1n}\\*
\nonumber&&+ q_\text{\text{L}}\tanh r_{\Omega}\proj{0\,n}{1\, n+1}\Big)+\frac{\sqrt{(n+1)(n+2)}}{\cosh^2r_{\Omega}}\\*
&&\times q_\text{\text{R}}\,q_\text{\text{L}}^*\tanh r_{\Omega}\proj{1\, n+2}{1n}+(\text{H.c.})_{_{\substack{\text{non-}\\\text{diag.}}}}.
\end{eqnarray}
 The eigenvalues of $\sigma_{AR}$ only depend on $|q_\text{R}|$ and $|q_\text{L}|$ and not on the relative phase between them. This means that the entanglement is insensitive to the election of this phase.

The two extreme cases when $q_R=1$ and $q_L=1$ are analytically solvable since the partial transpose density matrix has a block diagonal structure as it was shown in previous works \cite{Alicefalls}. However, for all other cases,  the matrix is no longer block diagonal and the eigenvalues of the partial transpose density matrix are computed numerically. The resulting negativity between Alice-Rob and Alice-AntiRob modes is plotted in Fig. (\ref{figbosons}) for different values of $|q_\text{\text{R}}|=1,0.9,0.8,0.7$.  $|q_\text{\text{R}}|=1$ corresponds to the canonical case studied in the literature \cite{Alicefalls}. 

\begin{figure}[h]
\begin{center}
\includegraphics[width=.50\textwidth]{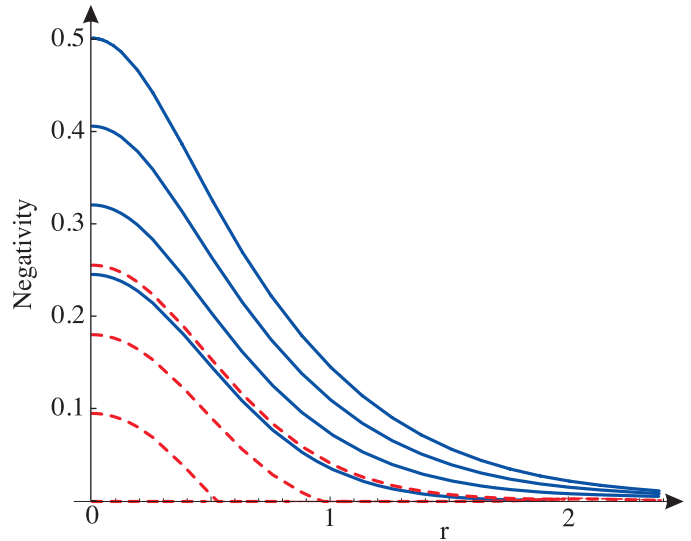}
\end{center}
\caption{Negativity for the bipartition Alice-Rob (Blue continuous) and Alice-AntiRob (red dashed) as a function of $r_{\Omega}=\operatorname{artanh}{e^{-\pi\Omega_a/a}}$ for various choices of $|q_\text{\text{R}}|$. The blue continuous (red dashed) curves from top to bottom (from bottom to top) correspond, to $|q_\text{\text{R}}|=1,0.9,0.8,0.7$  respectively .}
\label{figbosons}
\end{figure}
In the bosonic case, the entanglement between the Alice-Rob and Alice-AntiRob modes always vanishes in the infinite acceleration limit. Interestingly,  there is no fundamental difference in the degradation of entanglement  for different choices of $|q_\text{\text{R}}|$. The entanglement always degrades with acceleration at the same rate. There is no special Unruh state which degrades less with acceleration.

\section{Wave packets: recovering the single-mode approximation\label{sec:peaking}}

The entanglement analysis of section \ref{sec:entganglementrev}
assumes Alice's state
to be a Minkowski particle with a sharp Minkowski momentum and Rob's
state to be an Unruh particle with sharp Unruh frequency, 
such that Rob's linear combination of the two Unruh modes 
is specified by the two
complex-valued parameters $q_\text{\text{R}}$ and $q_\text{\text{L}}$ satisfying 
${|q_\text{\text{R}}|}^2 + {|q_\text{\text{L}}|}^2=1$. 
The Alice and Rob states are further
assumed to be orthogonal to each other, 
so that the system can be treated as
bipartite. We now discuss the sense in which these assumptions are a
good approximation to Alice and Rob states that can be built as
Minkowski wave packets.

Recall that a state with a sharp frequency, be it Minkowski or Unruh,
is not normalisable and should be understood as the idealisation of a
wave packet that contains a continuum of frequencies with an
appropriate peaking. Suppose that the Alice and Rob states are
initially set up as Minkowski wave packets, peaked about distinct
Minkowski momenta and having negligible overlap, so that the bipartite
assumption is a good approximation. The transformation between the
Minkowski and Unruh bases is an integral transform, given by 
\eqref{eq:Mink-v-Unruh}
and~\eqref{eq:alphas}: 
can the Rob state
be arranged to be peaked about a single Unruh frequency? If so, how
are the frequency uncertainties on the Minkowski and Unruh sides
related?

\subsection{Massless scalar field}\label{sec:peakingmassless}

We focus first on the massless scalar field of 
section~\ref{sec:entganglementrev}. 
The massive scalar field will be discussed 
in section~\ref{subsec:massivescalar}. 
We expect the analysis for fermions to be qualitatively
similar.

Consider a packet of Minkowski creation operators $a_{\omega,\text{M}}^\dag$
smeared with a weight function~$f(\omega)$. 
We wish to express this packet in terms of
Unruh creation operators $A_{\Omega,\text{\text{R}}}^\dag$ and $A_{\Omega,\text{\text{L}}}^\dag$ 
smeared with the weight
functions $g_{\text{R}}(\Omega)$ and $g_{\text{L}}(\Omega)$, so that 
\begin{align}
\int_{0}^{\infty}
&f(\omega) \, a_{\omega,\text{M}}^\dag \, d\omega
\notag
\\
& 
\hspace{2ex}
= 
\int_0^{\infty}
\left(
g_{\text{R}}(\Omega)A_{\Omega,\text{\text{R}}}^\dag+g_{\text{L}}(\Omega) A_{\Omega,\text{\text{L}}}^\dag
\right)
\text{d}\Omega. 
\label{eq:smeared-creator}
\end{align}
From \eqref{eq:Aaboth-transform} it follows that the smearing functions are related by
\begin{align}
g_\text{\text{R}}(\Omega) &= \int_{0}^{\infty}\alpha^\text{\text{R}}_{\omega\Omega}f(\omega) \, \text{d}\omega , \nonumber \\
g_\text{\text{R}}(\Omega) &= \int_{0}^{\infty}\alpha^\text{\text{R}}_{\omega\Omega}f(\omega) \, \text{d}\omega , \notag  \\
f(\omega) &= \int_{0}^{\infty}\left[(\alpha^\text{\text{R}}_{\omega\Omega})^{\ast}g_\text{\text{R}}(\Omega) 
+(\alpha^\text{\text{L}}_{\omega\Omega})^{\ast}g_\text{\text{L}}(\Omega)\right]\text{d}\Omega . 
\label{eq:f-gs-transform}
\end{align}
By~\eqref{eq:alphas}, equations \eqref{eq:f-gs-transform}
are recognised as a Fourier transform pair between the variable $\ln
(\omega\dimctwo) \in\BbbR$ on the Minkowski side and the variable
$\pm\Omega\in\BbbR$ on the Unruh side: the full real line on the Unruh
side has been broken into the Unruh frequency $\Omega \in\BbbR^+$ and
the discrete index R, L. 
All standard properties of
Fourier transforms thus apply. 
Parseval's theorem takes the form 
\begin{align}
\int_0^\infty {|f(\omega)|}^2 \, \text{d}\omega 
= 
\int_{0}^{\infty} 
\left(|g_\text{\text{R}}(\Omega)|^2+|g_\text{\text{L}}(\Omega)|^2 \right) 
\text{d}\Omega,
\end{align}
and the uncertainty 
relation reads 
\begin{align}
\label{eq:gen-uncertainty}
(\Delta \Omega)  
\bigl(\Delta \ln(\omega\dimctwo)\bigr) \ge \tfrac12, 
\end{align}
where $\Delta \Omega$ is understood by combining contributions from
$g_\text{\text{R}}(\Omega)$ and $g_\text{\text{L}}(\Omega)$ in the sense
of~\eqref{eq:f-gs-transform}.  Note that since equality in 
\eqref{eq:gen-uncertainty} holds only for Gaussians, any state in
which one of $g_\text{\text{R}}(\Omega)$ and ${}g_\text{\text{L}}(\Omega)$ vanishes will
satisfy \eqref{eq:gen-uncertainty} with a genuine inequality.

As a concrete example, with a view to optimising the peaking both in
Minkowski frequency and in
Unruh frequency, 
consider a Minkowski smearing function that is
a Gaussian in  $\ln(\omega\dimctwo)$, 
\begin{align}
\label{eq:gaussian-in-log}
f(\omega) = 
\left(\frac{\lambda}{\pi\omega^2}\right)^{\!\!1/4}
\exp\left\{- \tfrac12 \lambda
{\bigl[\ln(\omega/\omega_0)\bigr]}^2
\right\}
{(\omega/\omega_0)}^{-i\mu},
\end{align}
where $\omega_0$ and $\lambda$ are positive parameters and $\mu$ is a real-valued parameter. 
$\lambda$~and $\mu$ are dimensionless and $\omega_0$ has the dimension of inverse length. 
Note that $f$~is normalised, $\int_{0}^{\infty} {|f(\omega)|}^2 \, d\omega =1$. 
The expectation value and uncertainty of $\ln(\omega\dimctwo)$ are those of 
a standard Gaussian, $\langle \ln(\omega\dimctwo) \rangle = \ln(\omega_0\dimctwo)$ 
and $\Delta \ln(\omega\dimctwo) = {(2\lambda)}^{-1/2}$, 
while the expectation value and uncertainty of $\omega$ are given by 
\begin{eqnarray}
\langle \omega \rangle &=& \exp\left(\tfrac14\lambda^{-1}\right)\nonumber \\
\Delta\omega &=& \langle \omega \rangle {\left[\exp\bigl(\tfrac12\lambda^{-1}\bigr) -1 \right]}^{1/2}. 
\end{eqnarray}

The Unruh smearing functions are cropped Gaussians, 
\begin{align}
\label{eq:gs-gauss}
g_\text{\text{R}}(\Omega) &= \frac{1}{{(\pi\lambda)}^{1/4}}\exp\!\left[- \tfrac12 \lambda^{-1}{(\Omega - \epsilon\mu)}^2 \right] 
{(\omega_0\dimctwo)}^{i\epsilon\Omega},\nonumber \\
g_\text{\text{L}}(\Omega) &= \frac{1}{{(\pi\lambda)}^{1/4}}\exp\left[- \tfrac12 \lambda^{-1}{(\Omega + \epsilon\mu)}^2\right] 
{(\omega_0\dimctwo)}^{-i\epsilon\Omega}.
\end{align}
For $\epsilon\mu \gg \lambda^{1/2}$, $g_\text{\text{L}}(\Omega)$ is small and $g_\text{\text{R}}(\Omega)$ 
is peaked around $\Omega = \epsilon\mu$ with uncertainty ${(\lambda/2)}^{1/2}$; 
conversely, for $\epsilon\mu \ll -\lambda^{1/2}$, $g_\text{\text{R}}(\Omega)$ is small 
and $g_\text{\text{L}}(\Omega)$ is peaked around $\Omega = - \epsilon\mu$ with 
uncertainty ${(\lambda/2)}^{1/2}$. Note that in these limits, 
the relative magnitudes of $g_\text{\text{L}}(\Omega)$ and $g_\text{\text{R}}(\Omega)$ are 
consistent with the magnitude of the smeared mode Minkowski mode 
function $\int_{0}^{\infty}f(\omega) \, u_{\omega,\text{M}} \, d\omega$ 
in the corresponding regions of Minkowski space: 
a contour deformation argument shows that for $\epsilon\mu \gg \lambda^{1/2}$ 
the smeared mode function 
is large in the region $t+x>0$ and small in the region $t+x<0$, 
while for $\epsilon\mu \ll -\lambda^{1/2}$ it is large in the 
region $t-x>0$ and small in the region $t-x<0$. 

Now, let the Rob state have the smearing
function~\eqref{eq:gaussian-in-log}, and choose for Alice any state
that has negligible overlap with the Rob state, for example by taking
for Alice and Rob distinct values of~$\epsilon$. For $|\mu| \gg
\lambda^{1/2}$ and $\lambda$ not larger than of order unity, the
combined state is then well approximated by the single Unruh frequency
state of section \ref{sec:entganglementrev}
with $\Omega=|\mu|$ and with one of $q_\text{\text{R}}$ and
$q_\text{\text{L}}$ vanishing. In this case we hence recover the 
results in~\cite{Alicefalls}. 
To build a Rob state that is 
peaked about a single Unruh frequency with 
comparable $q_\text{\text{R}}$ and~$q_\text{\text{L}}$, 
so that the results of section 
\ref{sec:entganglementrev} are recovered, we may take 
a Minkowski smearing function that
is a linear combination of \eqref{eq:gaussian-in-log} and its complex
conjugate.

While the phase factor ${(\omega/\omega_0)}^{-i\mu}$ in the Minkowski
smearing function \eqref{eq:gaussian-in-log} is essential for
adjusting the locus of the peak in the Unruh smearing functions,
the choice of a logarithmic Gaussian for the magnitude appears not
essential. We have verified that similar results ensue with the
choices 
\begin{align}
f(\omega) = 
\frac{2^{\lambda}
{(\omega/\omega_0)}^{\lambda-i\mu} \exp(-\omega/\omega_0)}
{\sqrt{\omega\Gamma(2\lambda)}}
\end{align}
and 
\begin{align}
f(\omega) = 
\frac{{(\omega/\omega_0)}^{-i\mu}}{\sqrt{2\omega K_0(2\lambda)}}
\exp
\left[- \frac{\lambda}{2} 
\left(\frac{\omega}{\omega_0}+ \frac{\omega_0}{\omega}\right)
\right], 
\end{align}
for which the respective Unruh smearing functions can be expressed respectively in terms of
the gamma-function and a modified Bessel function. 

\subsection{Massive scalar field\label{subsec:massivescalar}}

For a scalar field of mass $m>0$, the 
Minkowski modes of the Klein-Gordon equation are
\begin{align}
\label{eq:massivescalar-Mmode}
u_{k,\text{M}} (t,x) = \frac{1}{\sqrt{4\pi \omega}}\exp(-i\omega t + i kx),
\end{align}
where $k\in\BbbR$ is the Minkowski momentum and $\omega \equiv \omega_k =\sqrt{m^2+k^2}$ is the Minkowski frequency. 
These modes are delta-normalised in $k$ as usual. 
The Rindler modes are \cite{Takagi}
\begin{eqnarray}
u_{\Omega,\text{I}} (t,x) &=& N_{\Omega}\exp\left[ -\frac{i\Omega}{2}\ln\left(\frac{x+t}{x-t}\right)\right],\nonumber \\
u_{\Omega,\text{II}} (t,x) &=& N_{\Omega}\exp\left[ -\frac{i\Omega}{2}\ln\left(\frac{-x+t}{-x-t}\right)\right],
\end{eqnarray}
where $N_{\Omega}= \frac{\sqrt{\sinh \pi\Omega}}{\pi}K_{i\Omega}\bigl(m \sqrt{x^2-t^2}\,\bigr)$ and 
$\Omega>0$ is the (dimensionless) Rindler frequency. These modes are delta-normalised in~$\Omega$. 
The Unruh modes $u_{\Omega,\text{\text{R}}}$ and $u_{\Omega,\text{\text{L}}}$ 
are as in~\eqref{eq:unruhmodes}. 
Note that in the Minkowski modes \eqref{eq:massivescalar-Mmode}
the distinction between 
the left-movers and the right-movers is in the sign of the label 
$k\in\BbbR$, but in the Rindler and Unruh modes the right-movers and the left-movers 
do not decouple, owing to the 
asymptotic behaviour of the solutions at the Rindler spatial infinity. 
The Rindler and Unruh modes do therefore not carry an index $\epsilon$ 
that would distinguish the right-movers and the left-movers. 

The transformation between the Minkowski and Unruh modes 
can be found by the methods of~\cite{Takagi}. 
In our notation, the transformation reads 
\begin{eqnarray}
\label{eq:alphas-massive}
u_{\Omega,\text{\text{R}}} &=& \int_{-\infty}^{\infty}(\alpha^\text{\text{R}}_{k\Omega})^{\ast}u_{k,\text{M}} \, \text{d}k , \nonumber \\
u_{\Omega,\text{\text{L}}} &=& \int_{-\infty}^{\infty}(\alpha^\text{\text{L}}_{k\Omega})^{\ast}u_{k,\text{M}} \, \text{d}k , \nonumber \\
u_{k ,\text{M}} &=& \int_{0}^{\infty}
\left(\alpha^\text{\text{R}}_{k\Omega} u_{\Omega,\text{\text{R}}} +\alpha^\text{\text{L}}_{k\Omega} u_{\Omega,\text{\text{L}}} \right)\text{d}\Omega, 
\end{eqnarray}
where
\begin{eqnarray}\label{eq:alphas-massive2}
\alpha^\text{\text{R}}_{k\Omega} &=& 
\frac{1}{\sqrt{2\pi\omega}}
{\left(\frac{\omega+k}{m}\right)}^{i\Omega} ,\nonumber  \\
\alpha^\text{\text{L}}_{k\Omega} &=&
\frac{1}{\sqrt{2\pi\omega}}
{\left(\frac{\omega+k}{m}\right)}^{-i\Omega}.
\end{eqnarray}
Transformations for the various operators read hence as 
in section \ref{sec:entganglementrev}
but with the replacements 
\begin{equation}
\label{eq:massless-to-massive}
\omega\to k, \quad
\int_0^\infty \text{d}\omega \longrightarrow \int_{-\infty}^\infty \text{d}k 
\end{equation}
and no $\epsilon$. In particular, 
\begin{eqnarray}
\label{eq:Aaboth-massive-transform}
a_{\Omega,\text{\text{R}}}&=&\int_{-\infty}^{\infty}  \alpha^\text{\text{R}}_{k\Omega} \, a_{k,\text{M}} \, \text{d}k \nonumber \\
a_{\Omega,\text{\text{L}}} &=& \int_{-\infty}^{\infty}  \alpha^\text{\text{L}}_{k\Omega} \, a_{k,\text{M}} \, \text{d}k  \\
a_{k,\text{M}} &=&  \int_{0}^{\infty}\left((\alpha^\text{\text{R}}_{k\Omega})^{\ast} A_{\Omega,\text{\text{R}}}+(\alpha^\text{\text{L}}_{k\Omega})^{\ast} A_{\Omega,\text{\text{L}}}\right)\text{d}\Omega.\nonumber
\end{eqnarray}

To consider peaking of Minkowski wave packets in the Unruh frequency, 
we note that the transform \eqref{eq:Aaboth-massive-transform} with
\eqref{eq:alphas-massive2} is now a Fourier transform between the
Minkowski rapidity $\tanh^{-1} (k/\omega) = \ln[(\omega+k)/m]\in\BbbR$
and $\pm\Omega\in\BbbR$. The bulk of the massless peaking discussion
of section \ref{sec:peakingmassless}
goes hence through with the replacements
\eqref{eq:massless-to-massive} and $\omega\dimctwo \to
(\omega+k)/m$. The main qualitative difference is that in the massive
case one cannot appeal to the decoupling of the right-movers and
left-movers when choosing for Alice and Rob states that have
negligible overlap.

\section{Unruh  entanglement degradation for Dirac fields}\label{sec5}

 Statistics plays a very important role in the behaviour of entanglement described by observers in uniform acceleration. While entanglement vanishes in the limit of infinite acceleration in the bosonic case \cite{Alicefalls,Edu4}, it remains finite for Dirac fields \cite{AlsingSchul,Edu2}. Therefore, it is interesting to revise the analysis of entanglement between Dirac fields for different elections of Unruh modes.
\subsection{Dirac fields}
In a parallel analysis to the bosonic case, we consider a Dirac field $\phi$ satisfying the
equation  $\{i\gamma^{\mu}(\partial_{\mu}-\Gamma_{\mu})+m\}\phi=0$
where $\gamma^{\mu}$ are the Dirac-Pauli matrices and $\Gamma_{\mu}$ are spinorial
affine connections. The field expansion in terms of the Minkowski solutions of the Dirac equation is
\begin{equation}
\phi=N_\text{M}\sum_k\left(c_{k,\text{M}}\, u^+_{k,\text{M}}+ d_{k,\text{M}}^\dagger \ u^-_{k,\text{M}}\right),
\end{equation}
Where $N_\text{M}$ is a normalisation constant and the label $\pm$  denotes respectively positive and negative energy solutions (particles/antiparticles) with respect to the Minkowskian Killing vector field $\partial_{t}$. The label $k$ is a multilabel including energy and spin $k=\{E_\omega,s\}$ where $s$ is the component of the spin on the quantisation direction. $c_k$ and $d_k$ are the particle/antiparticle operators that satisfy the usual anticommutation rule
\begin{equation}
\{c_{k,\text{M}},c_{k',\text{M}}^\dagger\}=\{ d_{k,\text{M}},d_{k',\text{M}}^\dagger\}=\delta_{kk'},
\end{equation}
and all other anticommutators vanishing. The Dirac field operator in terms of Rindler modes is given by
\begin{equation}
\phi\!=N_\text{\text{R}}\sum_j\left(c_{j,\text{I}} u^+_{j,\text{I}}+ d_{j,\text{I}}^\dagger u^-_{j,\text{I}}+c_{j,\text{II}} u^+_{j,\text{II}}+ d_{j,\text{II}}^\dagger  u^-_{j,\text{II}}\!\right),
\end{equation}
Where $N_\text{R}$ is, again, a normalisation constant.  $c_{j,\Sigma},d_{j,\Sigma}$ with $\Sigma=\text{I},\text{II}$ represent Rindler particle/antiparticle operators. The usual anticommutation rules again apply. Note that operators in different regions $\Sigma=\text{I},\text{II}$ do not commute but anticommute. $j=\{E_\Omega,s'\}$ is again a multi-label including all the degrees of freedom. Here  $u^\pm_{k,\text{I}}$ and $u^\pm_{k,\text{II}}$  are the positive/negative frequency solutions of the Dirac equation in Rindler coordinates with respect to the Rindler timelike Killing vector field in region $\text{I}$ and $\text{II}$, respectively. The modes  $u^\pm_{k,\text{I}}$, $u^\pm_{k,\text{II}}$  do not have support outside the right, left Rindler wedge. 
The annihilation operators $c_{k,\text{M}},d_{k,\text{M}} $ define the Minkowski vacuum $\ket{0}_\text{M}$ which must satisfy
\begin{equation}
c_{k,\text{M}}\ket{0}_\text{M}= d_{k,\text{M}}\ket{0}_\text{M}=0, \qquad \forall k.
\end{equation}
In the same fashion $c_{j,\Sigma},d_{j,\Sigma}$, define the Rindler vacua in regions $\Sigma=\text{I},\text{II}$
\begin{equation}
c_{j,\text{\text{R}}}\ket{0}_\Sigma=d_{j,\text{\text{R}}}\ket{0}_\Sigma=0, \qquad \forall j, \, \Sigma=\text{I},\text{II}.
\end{equation}	
The transformation  between the Minkowski and Rindler modes is given by
\begin{eqnarray}
\nonumber u^+_{j,\text{M}}&=&\sum_k\left[\alpha^\text{I}_{jk}u^+_{k,\text{I}} + \beta^{\text{I}*}_{jk} u^-_{k,\text{I}}+\alpha^\text{II}_{jk}u^+_{k,\text{II}}\right.\\*
&& +\left. \beta^{\text{II}*}_{jk} u^-_{k,\text{II}}\right].
\end{eqnarray}
The coefficients which relate both set of modes are given by the inner product
\begin{equation}
(u_k,u_j)=\int d^3x\, u_k^\dagger u_j,
\end{equation}
so that he  Bogoliubov coefficients
are, after some elementary but lengthly algebra \cite{Jauregui,Langlois},
\begin{equation}
\alpha^\text{I}_{jk}=e^{i\theta E_\Omega}\frac{1+i}{2\sqrt{\pi E_\omega}}\,\frac{e^{\pi E_\Omega/2}}{\sqrt{e^{\pi E_\Omega}+e^{-\pi E_\Omega}}}\delta_{ss'},
\end{equation}
\begin{equation}
\beta^\text{I}_{jk}=-e^{i\theta E_\Omega}\frac{1+i}{2\sqrt{\pi E_\omega}}\,\frac{e^{-\pi E_\Omega/2}}{\sqrt{e^{\pi E_\Omega}+e^{-\pi E_\Omega}}}\delta_{s s'},
\end{equation}
where $E_\Omega$ is the energy of the Rindler mode $k$, $E_\omega$ is the energy of the Minkowski mode $j$ and $\theta$ is a parameter defined such that it satisfies the condition $E_\Omega=m\cosh\theta$ and $|\bm k_\Omega|=m\sinh\theta$   (see \cite{Jauregui}).  One can verify that $\alpha^\text{II}=(\alpha^\text{I})^*$ and $\beta^\text{II}=(\beta^\text{I})^*$.  Defining $\tan r_{\Omega}=e^{-\pi E_\Omega}$
the coefficients become
\begin{eqnarray}
\alpha^\text{I}_{jk}=e^{i\theta E_\Omega}\frac{1+i}{2\sqrt{\pi E_\omega}}\,\cos r_{\Omega}\,\delta_{s s'},\nonumber\\
\beta^\text{I}_{jk}=-e^{i\theta E_\Omega}\frac{1+i}{2\sqrt{\pi E_\omega}}\,\sin r_{\Omega}\,\delta_{s s'}.
\end{eqnarray}
Finally, taking into account that $c_{j,\text{M}}=(u^+_{j,\text{M}},\phi)$  we find the Minkowski particle annihilation operator to be
\begin{equation}
 c_{j,\text{M}}=\sum_{k}\left[\alpha^{\text{I}*}_{jk}c_{k,\text{I}} + \beta^\text{I}_{jk} d_{k,s,\text{I}}^\dagger+\alpha^{\text{II}*}_{jk}c_{k,\text{II}}+\beta^\text{II}_{jk}d_{k,\text{II}}^{\dagger}\right].
\end{equation}

We now consider the transformations between states in different basis. For this we define an arbitrary element of the Dirac field Fock basis for each mode as
\begin{equation}
\ket{F_k}=\ket{F_k}_\text{\text{R}}\otimes \ket{F_k}_\text{\text{L}},
\end{equation}
where 
\begin{eqnarray}
\nonumber\ket{F_k}_\text{R}&=&\ket{n}^+_\text{I}\ket{m}^-_\text{II},\\*
\ket{F_k}_\text{L}&=&\ket{p}^-_\text{I}\ket{q}^+_\text{II}.
\end{eqnarray}
Here the $\pm$  indicates particle/antiparticle. Operating with the Bogoliubov coefficients making this tensor product structure explicit we obtain
\begin{equation}\label{annihil}
c_{j,\text{M}}=N_j\sum_k\Bigg[\chi^*(C_{k,\text{\text{R}}}\otimes\openone_\text{\text{L}})+\chi(\openone_\text{\text{R}}\otimes C_{k,\text{\text{L}}})\Bigg],
\end{equation}
where
\begin{equation}\label{ene}
N_j=\frac{1}{2\sqrt{\pi E_\omega}}\qquad  \chi=(1+i)e^{i\theta E_{\Omega}},
\end{equation}
and the operators
\begin{eqnarray}\label{Unruhop}
\nonumber C_{k,\text{\text{R}}}&\equiv&\left(\cos r_k\, c_{k,\text{I}}-\sin r_k\, d^\dagger_{k,\text{II}}\right),\\*
C_{k,\text{\text{L}}}&\equiv&\left(\cos r_k\, c_{k,\text{II}}-\sin r_k\, d^\dagger_{k,\text{I}}\right)
\end{eqnarray}
are the so-called Unruh operators. 

It can be shown \cite{ch1} that for a massless Dirac field the Unruh operators have the same form as Eq. \eqref{Unruhop} however in this case $\tan r_{k}=e^{-\pi\Omega_a/a}$.

In the massless case, to find the Minkowski vacuum in the Rindler basis we consider the following ansatz
\begin{equation}
\ket{0}_\text{M}=\bigotimes_{\Omega}\ket{0_\Omega}_\text{M},
\end{equation}
where $\ket{0_\Omega}_\text{M}=\ket{0_\Omega}_\text{\text{R}}\otimes\ket{0_\Omega}_\text{\text{L}}$.
We find that
\begin{eqnarray}
\label{vauno}
\ket{0_{\Omega}}_\text{\text{R}}&=&\sum_{n,s}\left(F_{n,\Omega,s}\ket{n_{\Omega,s}}^+_\text{I}\ket{n_{\Omega,-s}}^-_\text{II}\right)\nonumber \\
\ket{0_{\Omega}}_\text{\text{L}}&=&\sum_{n,s}\left(G_{n,\Omega,s}\ket{n_{\Omega,s}}^-_\text{I}\ket{n_{\Omega,-s}}^+_\text{II}\right),
\end{eqnarray}
where the label $\pm$ denotes particle/antiparticle modes  and $s$ labels the spin. The minus signs on the spin label in region $\text{II}$ show explicitly that spin, as all the magnitudes which change under time reversal, is opposite in region I with respect to region II.

We obtain the form of the coefficients $F_{n,\Omega,s},G_{n,\Omega,s}$ for the vacuum by imposing that the Minkowski vacuum is annihilated by the particle annihilator for all frequencies and values for the spin third component.

\subsection{Grassman scalars}

Since the simplest case that preserves the fundamental Dirac characteristics corresponds to Grassman scalars, we study them in what follows. Moreover,  the entanglement in non-inertial frames between scalar fermionic fields has been extensively studied under the single mode approximation in the literature \cite{AlsingSchul}. In this case, the Pauli exclusion principle limits the sums \eqref{vauno} and only the two following terms contribute
\begin{eqnarray}
\label{vaunos}
\ket{0_\Omega}_\text{\text{R}}&=&F_0\ket{0_\Omega}^+_\text{I}\ket{0_\Omega}^-_\text{II}+F_1\ket{1_\Omega}^+_\text{I}\ket{1_\Omega}^-_\text{II},\nonumber \\
\ket{0_\Omega}_\text{\text{L}}&=&G_0\ket{0_\Omega}^-_\text{I}\ket{0_\Omega}^+_\text{II}+G_1\ket{1_\Omega}^-_\text{I}\ket{1_\Omega}^+_\text{II}.
\end{eqnarray}
Due to the anticommutation relations we must introduce the following sign conventions
\begin{eqnarray}
\ket{1_\Omega}^+_\text{I}\!\ket{1_\Omega}^-_\text{II}&=&d^\dagger_{\Omega,\text{II}}c^\dagger_{\Omega,\text{I}}\!\ket{0_\Omega}^+_\text{I}\!\ket{0_\Omega}^-_\text{II},\nonumber \\
&=&-c^\dagger_{\Omega,\text{I}}d^\dagger_{\Omega,\text{II}}\!\ket{0_\Omega}^+_\text{I}\!\ket{0_\Omega}^-_\text{II},\nonumber\\*
\nonumber\ket{1_\Omega}^-_\text{I}\!\ket{1_\Omega}^+_\text{II}&=&c^\dagger_{\Omega,\text{II}}d^\dagger_{\Omega,\text{I}}\!\ket{0_\Omega}^-_\text{I}\!\ket{0_\Omega}^+_\text{II},\nonumber \\
&=&-d^\dagger_{\Omega,\text{I}}c^\dagger_{\Omega,\text{II}}\!\ket{0_\Omega}^-_\text{I}\!\ket{0_\Omega}^+_\text{II}.
\end{eqnarray}
We obtain the form of the coefficients by imposing that $c_{\omega,\text{M}}\ket{0_\Omega}_\text{M}=0$ which translates into $C_{\Omega,\text{\text{R}}}\ket{0_\Omega}_\text{\text{R}}=C_{\Omega,\text{\text{L}}}\ket{0_\Omega}_\text{\text{L}}=0$. Therefore
\begin{eqnarray}
\label{co1} C_{\Omega,\text{\text{R}}}\left(F_0\ket{0_\Omega}^+_\text{I}\ket{0_\Omega}^-_\text{II}+F_1\ket{1_\Omega}^+_\text{I}\ket{1_\Omega}^-_\text{II}\right)&=&0,\nonumber \\
\label{co2} C_{\Omega,\text{\text{L}}}\left(G_0\ket{0_\Omega}^-_\text{I}\ket{0_\Omega}^+_\text{II}+G_1\ket{1_\Omega}^-_\text{I}\ket{1_\Omega}^+_\text{II}\right)&=&0.
\end{eqnarray}
These conditions imply that
\begin{eqnarray}
F_1\cos r_{\Omega}&-&F_0\sin r_{\Omega}=0\Rightarrow F_1=F_0\tan r_{\Omega},\nonumber \\
G_1\cos r_{\Omega}&+&G_0\sin r_{\Omega}=0\Rightarrow G_1=-G_0\tan r_{\Omega},
\end{eqnarray}
which together with the normalisation conditions $\bra{0_\Omega}_\text{\text{R}}\ket{0_\Omega}_\text{\text{R}}=1$ and $\bra{0_\Omega}_\text{\text{L}}\ket{0_\Omega}_\text{\text{L}}=1$ yield
\begin{eqnarray}
F_0&=&\cos r_{\Omega},\qquad F_1=\sin r_{\Omega},\\
G_0&=&\cos r_{\Omega}\qquad G_1=-\sin r_{\Omega}.\nonumber
\end{eqnarray}
Therefore  the vacuum state is given by,
 \begin{align}\label{vacgrassman} 
\nonumber\ket{0_{\Omega}}&=\left(\cos r_{\Omega}\ket{0_{\Omega}}^+_\text{I}\ket{0_{\Omega}}^-_\text{II}+\sin r_{\Omega}\ket{1_{\Omega}}^+_\text{I}\ket{1_{\Omega}}^-_\text{II}\right)\\
&\otimes\left(\cos r_{\Omega}\ket{0_{\Omega}}^-_\text{I}\ket{0_{\Omega}}^+_\text{II}-\sin r_{\Omega}\ket{1_{\Omega}}^-_\text{I}\ket{1_{\Omega}}^+_\text{II}\right)\!,
\end{align}
which is compatible with the result obtained with the Unruh modes. For convenience, we introduce the following notation, \begin{equation}\label{shortnot}
\ket{n n' n'' n'''}_{\Omega}\equiv\ket{n_{\Omega}}^+_\text{I}\ket{n'_{\Omega}}^-_\text{II}\ket{n''_{\Omega}}^-_\text{I}\ket{n'''_{\Omega}}^+_\text{II},
\end{equation}
in which the vacuum state is written as,
\begin{eqnarray}\label{shortvac}
\ket{0_{\Omega}}&=&\cos^2 r_{\Omega}\ket{0000}_{\Omega}-\sin r_{\Omega}\cos r_{\Omega}\ket{0011}_{\Omega}\\
&+&\sin r_{\Omega}\cos r_{\Omega} \ket{1100}_{\Omega}-\sin^2 r_{\Omega} \ket{1111}_{\Omega}.\nonumber
\end{eqnarray}

The Minkowskian one particle state is obtained by applying the creation operator to the vacuum state $\ket{1_{j}}_{\text{U}}=c_{\Omega,\text{U}}^\dagger\ket{0}_\text{M}$,
where the Unruh particle creator is a combination of the two Unruh operators $C_{\Omega,\text{\text{R}}}^\dagger$ and $C_{\Omega,\text{\text{L}}}^\dagger$,
\begin{equation}\label{creat}
c_{k,\text{U}}^\dagger\!=\!q_\text{\text{R}}(C^\dagger_{\Omega,\text{\text{R}}}\otimes\openone_\text{\text{L}})+q_\text{\text{L}}(\openone_\text{\text{R}}\otimes C^\dagger_{\Omega,\text{\text{L}}}).
\end{equation}
Since
\begin{eqnarray}
C^\dagger_{\Omega,\text{\text{R}}}&\equiv&\left(\cos r_{\Omega}\, c^\dagger_{\Omega,\text{I}}-\sin r_{\Omega}\, d_{\Omega,\text{II}}\right),\\*
C^\dagger_{\Omega,\text{\text{L}}}&\equiv&\left(\cos r_{\Omega}\, c^\dagger_{\Omega,\text{II}}-\sin r_{\Omega}\, d_{\Omega,\text{I}}\right),\nonumber
\end{eqnarray}
with $q_\text{\text{R}},q_\text{\text{L}}$ complex numbers satisfying $|q_\text{\text{R}}|^2+|q_\text{\text{L}}|^2=1$,
we obtain,
\begin{eqnarray}
\ket{1_{\Omega}}^+_\text{\text{R}}&=&C^\dagger_{\Omega,\text{\text{R}}}\ket{0_{\Omega}}_\text{\text{R}}=\ket{1_{\Omega}}^+_\text{I}\ket{0_{\Omega}}^-_\text{II}\\*
\ket{1_{\Omega}}^+_\text{\text{L}}&=&C^\dagger_{\Omega,\text{\text{L}}}\ket{0_{\Omega}}_\text{\text{L}}=\ket{0_{\Omega}}^-_\text{I}\ket{1_{\Omega}}^+_\text{II}\nonumber
\end{eqnarray}
and therefore,
\begin{equation}
\ket{1_{k}}^+_\text{U}=q_\text{\text{R}}\ket{1_{\Omega}}_\text{\text{R}}\otimes\ket{0_{\Omega}}_\text{\text{L}}+q_\text{\text{L}}\ket{0_{\Omega}}_\text{\text{R}}\otimes\ket{1_{\Omega}}_\text{\text{L}}.
\end{equation}
In the short notation we have introduced the state reads,  
\begin{eqnarray}\label{onegrassman}
\ket{1_{k}}^+_\text{U}&=& q_\text{\text{R}}\left[\cos r_{k}\ket{1000}_{\Omega}-\sin r_{\Omega}\ket{1011}_{\Omega}\right]\\*
&+&q_\text{\text{L}}\left[\sin r_{\Omega}\ket{1101}_{\Omega}+\cos r_{\Omega}\ket{0001}_{\Omega}\right].\nonumber
\end{eqnarray}

\subsection{Fermionic entanglement beyond the single mode approximation}

 Let us now consider the following fermionic maximally entangled state
\begin{equation}\label{eq:states}
\ket{\Psi}=\frac{1}{\sqrt2}\left(\ket{0_\omega}_{\text M}\ket{0_\Omega}_\text{U}+\ket{1_\omega}^+_{\text{M}}\ket{1_\Omega}^+_\text{U}\right),
\end{equation}
which is the fermionic analog to \eqref{maxent} and where the modes labeled with $\text{U}$ are Grassman Unruh modes. To  compute Alice-Rob partial density matrix  we trace over region II in in $\proj{\Psi}{\Psi}$ and obtain,  
\begin{eqnarray}
\nonumber\rho_{AR}\!\!&=&\!\!\frac{1}{2}\Big[C^4\proj{000}{000}+S^2C^2(\proj{010}{010}+\proj{001}{001})\\
\nonumber&&\!\!\!\!\!\!\!\!\!\!\!\!\!\!\!\!\!+\phantom{\Big[}S^4\proj{011}{011}+|q_\text{\text{R}}|^2(C^2\proj{110}{110}+S^2\proj{111}{111})\\*
\nonumber&&\!\!\!\!\!\!\!\!\!\!\!\!\!\!\!\!\!+\phantom{\Big[}\!\!|q_\text{\text{L}}|^2(S^2\!\proj{110}{110}\!+\!C^2\!\proj{100}{100})\!+\!q_\text{\text{R}}^*(C^3\!\proj{000}{110}\\*
\nonumber&&\!\!\!\!\!\!\!\!\!\!\!\!\!\!\!\!\!+S^2C\!\proj{001}{111})-q_\text{\text{L}}^*(C^2S\!\proj{001}{100}+S^3\!\proj{011}{110})\\*
&&\!\!\!\!\!\!\!\!\!\!\!\!\!\!\!\!\!-q_\text{R} q_\text{L}^*SC\!\proj{111}{100}\Big]+(\text{H.c.})_{_{\substack{\text{non-}\text{diag.}}}}
\end{eqnarray}
in the basis were $C=\cos r_{\Omega}$ and $S=\sin r_{\Omega}$. To compute the negativity, we first obtain the partial transpose density matrix (transpose only in the subspace of Alice or Rob) and find its negative eigenvalues. The partial transpose matrix is block diagonal and only the following two blocks contribute to negativity,
\begin{itemize}
\item $\{\ket{100},\ket{010},\ket{111}\}$
\end{itemize}
\begin{equation}
\frac12\left(\begin{array}{ccc}
 C^2|q_\text{\text{L}}|^2& C^3q_\text{\text{R}}^*&-q_\text{R}^* q_\text{L}SC\\
C^3q_\text{\text{R}}& S^2C^2& -q_\text{\text{L}} S^3\\
-q_\text{R} q_\text{L}^*SC & -q_\text{\text{L}}^* S^3&|q_\text{\text{R}}|^2S^2\\
\end{array}\right),
\end{equation}
\begin{itemize}
\item $\{\ket{000},\ket{101},\ket{011}\}$
\end{itemize}
\begin{equation}
\frac12\left(\begin{array}{ccc}
C^4&-q_\text{\text{L}}C^2S&0\\
-q_\text{\text{L}}^*C^2S& 0 & q_\text{\text{R}}^*S^2C \\
0& q_\text{\text{R}}S^2C & S^4 \\
\end{array}\right).
\end{equation}
were the basis used is $\ket{ijk}=\ket{i}_\text{M}\overbrace{\ket{j}_\text{I}^+\ket{k}^-_\text{I}}^{\text{Rob}}$.  Notice that although the system is bipartite, the dimension of the partial Hilbert space for Alice is lower than the dimension of the Hilbert space for Rob, which includes particle and antiparticle modes.
The eigenvalues only depend on $|q_\text{\text{R}}|$ and not on the relative phase between $q_\text{\text{R}}$ and $q_\text{\text{L}}$.

The density matrix for the Alice-AntiRob modes is obtained by tracing over region $\text{I}$ in $\proj{\Psi}{\Psi}$, \begin{eqnarray} 
\nonumber\rho_{A\bar R}\!\!&=&\!\!\frac{1}{2}\Big[C^4\proj{000}{000}+S^2C^2(\proj{010}{010}+\proj{001}{001})\\*
\nonumber&&\!\!\!\!\!\!\!\!\!\!\!\!\!\!\!\!\!+\phantom{\Big[}S^4\proj{011}{011}+|q_\text{\text{R}}|^2(C^2\proj{100}{100}+S^2\proj{101}{101})\\*
\nonumber&&\!\!\!\!\!\!\!\!\!\!\!\!\!\!\!\!\!+\!\!\phantom{\Big[}|q_\text{\text{L}}|^2(S^2\!\proj{111}{111}\!+\!C^2\!\proj{101}{101})\!+\!q_\text{\text{L}}^*(C^3\!\proj{000}{101}\\*
\nonumber&&\!\!\!\!\!\!\!\!\!\!\!\!\!\!\!\!\!+S^2C\!\proj{010}{111})+q_\text{\text{R}}^*(C^2S\!\proj{010}{100}+S^3\!\proj{011}{101})\\*
&&\!\!\!\!\!\!\!\!\!\!\!\!\!\!\!\!\!+q_\text{R} q_\text{L}^*SC\!\proj{100}{111}\Big]+(\text{H.c.})_{_{\substack{\text{non-}\\\text{diag.}}}}
\end{eqnarray}
In this case the blocks of the partial transpose density matrix which contribute to the negativity are, 
\begin{itemize}
\item $\{\ket{111},\ket{001},\ket{100}\}$
\end{itemize}
\begin{equation}
\frac12\left(\begin{array}{ccc}
 S^2|q_\text{\text{L}}|^2& S^3q_\text{\text{R}}^*&q_\text{R}^* q_\text{L}SC\\
S^3q_\text{\text{R}}& C^2S^2& q_\text{\text{L}} C^3\\
q_\text{R} q_\text{L}^*SC & q_\text{\text{L}}^* C^3&|q_\text{\text{R}}|^2C^2\\
\end{array}\right),
\end{equation}
\begin{itemize}
\item $\{\ket{011},\ket{110},\ket{000}\}$
\end{itemize}
\begin{equation}
\frac12\left(\begin{array}{ccc}
S^4&q_\text{\text{L}}S^2C&0\\
q_\text{\text{L}}^*S^2C& 0 & q_\text{\text{R}}^*C^2S \\
0& q_\text{\text{R}}C^2S & C^4 \\
\end{array}\right),
\end{equation}
where we have considered the basis $\ket{ijk}=\ket{i}_\text{M}\overbrace{\ket{j}_\text{II}^-\ket{k}^+_\text{II}}^{\text{Anti-Rob}}$. Once more, the eigenvalues only depend on $|q_\text{\text{R}}|$ and not on the relative phase between $q_\text{\text{R}}$ and $q_\text{\text{L}}$.

\begin{figure}[h]
\begin{center}
\includegraphics[width=.50\textwidth]{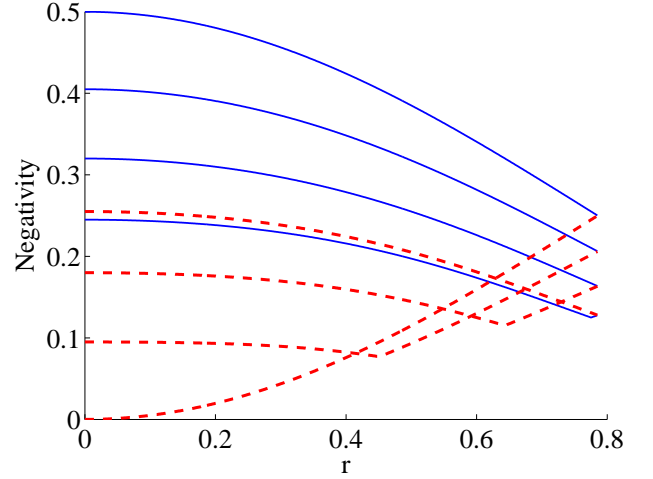}
\end{center}
\caption{Negativity for the bipartition Alice-Rob (Blue continuous) and Alice-AntiRob (red dashed) as a function of $r_{\Omega}=\arctan{e^{-\pi\Omega_a/a}}$ for various choices of $|q_\text{\text{R}}|$. The blue continuous (red dashed) curves from top to bottom (from bottom to top) correspond, to $|q_\text{\text{R}}|=1,0.9,0.8,0.7$  respectively. All the curves for Alice-AntiRob entanglement have a minimum, except from the extreme case $|q_R|=1$}
\label{bundlef}
\end{figure}

In Fig.~\ref{bundlef} we plot the entanglement between  Alice-Rob (solid line) and Alice-AntiRob (dashed line) modes quantified by the negativity as a function of acceleration for different choices of $|q_\text{\text{R}}|$ (in the range \mbox{$1\ge|q_\text{\text{R}}|>1/\sqrt{2}$}).   

We confirm that the case $|q_\text{\text{R}}|=1$ reproduced the results reported in the literature \cite{AlsingSchul}. The entanglement between Alice-Rob modes is degraded as the acceleration parameter increases reaching a non-vanishing minimum value in the infinite acceleration limit $a\rightarrow\infty$. However,  while the entanglement Alice-Rob decreases, entanglement between the  Alice-AntiRob partition (dashed line) grows.  Interestingly, the quantum correlations between the bipartitions Alice-Rob and Alice-AntiRob fulfill a conservation law $\mathcal{N}(\text{Alice-Rob})+\mathcal{N}(\text{Alice-AntiRob})=1/2$. 
Note that the choice $|q_\text{\text{R}}|=0$ corresponds to an exchange of the Alice-Rob and Alice-AntiRob bipartitions.  In such case, the entanglement between Alice and Anti-Robs's modes degrades with acceleration while the entanglement between Alice and Rob's modes grows.  In fact, regarding entanglement, the role of the Alice-Rob and Alice-AntiRob partitions are exchanged when $|q_\text{\text{R}}|<|q_\text{\text{L}}|$. This is because there is an explicit symmetry between field excitations in the Rindler wedges.  Therefore, we will limit our analysis to  $|q_\text{\text{R}}|>|q_\text{\text{L}}|$.

 In the fermionic case different choices of $|q_\text{\text{R}}|$ result in different degrees of entanglement between modes. In particular, the amount of entanglement in the limit of infinite acceleration depends on this choice. Therefore, we can find a special Unruh state which is more resilient  to entanglement degradation. The total entanglement is maximal in the infinite acceleration limit in the case $|q_\text{\text{R}}|=1$ (or $|q_\text{\text{L}}|=1$) in which  $\mathcal{N}_{\infty}(\text{Alice-Rob})=\mathcal{N}_{\infty}(\text{Alice-AntiRob})=0.25$.  In this case, the entanglement lost between Alice-Rob modes is completely compensated by the creation of entanglement between Alice-AntiRob modes.

In the case $|q_\text{\text{R}}|=|q_\text{\text{L}}|=1/\sqrt2$ we see that the behaviour of both bipartitions is identical. The entanglement from the inertial perspective is equally distributed between between the Alice-Bob and Alice-AntiBob partitions and adds up to
$\mathcal{N}(\text{Alice-Bob})+\mathcal{N}(\text{Alice-AntiBob}) = 0.5$ which corresponds to the total entanglement
between Alice-Bob when $|q_\text{\text{R}}|=1$. In the infinite acceleration limit, the case $|q_\text{\text{R}}|=|q_\text{\text{L}}|$ reaches the minimum total entanglement.   To understand this we note that the entanglement in the Alice-Rob bipartition for  $|q_\text{\text{R}}|>|q_\text{\text{L}}|$ is always monotonic. However, this is not the case for the entanglement between the Alice-AntiRob modes.  Consider the plot in Fig. \ref{bundlef} for the cases $|q_\text{R}|<1$,  for small accelerations, entanglement is degraded in both bipartitons.  However, as the acceleration increases, entanglement between Alice-AntiRob modes is created compensating the entanglement lost between Alice-Rob. The equilibrium point between degradation and creation  is the minimum  that Alice-AntiRob entanglement curves present. Therefore, if $|q_R|<1$ the entanglement lost is not entirely compensated by the creation of entanglement between Alice-AntiRob resulting in a less entangled state in the infinite acceleration limit. 

 In the case $|q_\text{\text{R}}|=|q_\text{\text{L}}|=1/\sqrt2$ entanglement is always degraded between Alice-AntiRob modes resulting in the state, among all the possible elections of Unruh modes,  with less entanglement in the infinite acceleration limit.

\section{Conclusions}\label{conclusions}

We have shown that the single-mode approximation used in the relativistic quantum information literature, especially to analyse entanglement between field modes from the perspective of observers in uniform acceleration, does not hold for general states. The single-mode approximation attempts to relate a single Minkowski frequency mode (observed by inertial observers) with a single Rindler frequency mode (observed by uniformly accelerated observers).

 We show that the state canonically analysed in the literature corresponds to a maximally entangled state of a Minkowski mode and a specific kind of Unruh mode (not two Minkowski modes). We analyse the entanglement between two bosonic or fermionic modes in the case when, from the inertial perspective, the state corresponds to a maximally entangled state between a Minkowski frequency mode and an arbitrary Unruh frequency mode.
 
We find that the entanglement between modes described by an Unruh observer and a Rindler observer constrained to move in Rindler region I (Alice-Rob) are always degraded with acceleration.  In the bosonic case, the entanglement between the inertial modes  and region II Rindler  modes (Alice-AntiRob) are also degraded with acceleration.  We find that, in this case,  the rate of entanglement degradation is independent of the choice of Unruh modes.

 For the fermionic case the entanglement between the inertial and region I Rindler modes (Alice-Rob) is degraded as the acceleration increases reaching a minimum value when it tends to infinity. There is, therefore, entanglement survival in the limit of infinite acceleration for any choice of Unruh modes. However, we find an important difference with the bosonic case: the amount of surviving entanglement depends on the specific election of such modes.

We also find that the entanglement between inertial and region II Rindler modes (Alice-AntiRob) can be  created and degraded depending on the election of Unruh modes. This gives rise to different values of entanglement in the infinite acceleration limit.  Interestingly,  in the fermionic case one can find a state which is most resilient to entanglement degradation. This corresponds to the Unruh mode with  $|q_\text{\text{R}}|=1$ which is the Unruh mode considered in the canonical studies of entanglement \cite{Alsingtelep,Alicefalls,AlsingSchul,Edu2,Edu3,Edu4}. It  could be argued that this is the most natural choice of Unruh modes since for this choice ($|q_R|=1$) the entanglement for very small accelerations ($a\rightarrow0$) is mainly  contained in the subsystem Alice-Rob. In this case, there is nearly no entanglement between the Alice-AntiRob modes.  However, other choices of Unruh modes become relevant if one wishes to consider an entangled state from the inertial perspective which involves only Minkowski frequencies. We have shown that a Minkowski wavepacket involving a superposition of general Unruh modes can  be constructed in such way that the corresponding Rindler state involves a single frequency.  This result is especially interesting since it presents an instance where the single-mode approximation can be considered recovering the standard results in the literature. 

\section{Acknowledgements}
We would like to thank R.~B. Mann, J.~Leon, B.~L Hu, P.~Alsing, T.~Ralph,  
T.~Downes and K. Pachucki for interesting discussions and useful comments.  
I. F was supported by EPSRC [CAF Grant EP/G00496X/2]. 
J.~L. was was supported in part by STFC (UK) Rolling Grant 
PP/D507358/1. 
E. M-M was supported by a CSIC JAE-PREDOC2007 Grant and by the Spanish MICINN Project FIS2008-05705/FIS.

\bibliographystyle{pr}

\end{document}